\documentclass{ckm}                 % twocolumn proceedings style
\usepackage{txfonts}            % if you have it, to get Times family
\usepackage{mathptmx}          % use the mathptmx package
\usepackage{rotate}
\usepackage{epsfig}
\usepackage{graphicx}% Include figure files
\usepackage{dcolumn}% Align table columns on decimal point
\usepackage{bm}% bold math
\confname{Workshop on the CKM Unitarity Triangle, IPPP Durham, April
  2003}

\title{CP Violation and Nonleptonic B-meson Decays} 

\author{Y.-Y. Keum and A. I. Sanda}

\address{Department of Physics Nagoya University}

\begin{document}

\begin{abstract}We discuss the perturbative QCD approach for exclusive
non-leptonic two body B-meson decays.
We briefly review its ingredients and some important theoretical
issues related to the factorization approach. 
PQCD results are compatible with present experimental data 
for the charmless B-meson decays. 
We predict the possibility of large direct CP
violation effects in $B^0 \to \pi^{+}\pi^{-}$ $(23\pm7 \%)$ and
$B^0\to K^{+}\pi^{-}$ $(-17\pm5\%)$. 
For charmed decay $B \to D^{(*)}\pi$, we get large non-factorizable
contributions for Color-suppressed decays and obtain  
$|a_2/a_1|\sim 0.4-0.5$ and $Arg(a_2/a_1)\sim -42^{o}$.
In the last section we investigate the method 
to extract the weak phases $\phi_2$ 
from $B \to \pi\pi$ decay and $\phi_3$ from $K\pi$ modes. 
From BaBar measurement of CP asymmetry 
for the $\pi^{+}\pi^{-}$ decay, the prefered CKM weak phases are:
$\phi_1=(24\pm2)^{o}$, $\phi_2=(78\pm 22)^{o}$ and $\phi_3=(78\pm 22)^{o}$. 
\end{abstract}

%% \maketitle needs to be after the author and address info and the
%% abstract... 
\maketitle

%% standard LaTeX from here on...

%%%%%%%%%%%%%%%%%%%%%% Definitions %%%%%%%%%%%%%%%%%%%%%%
\def \app{A_{\pi \pi}}
\def \bb{\overline{{\cal B}}}
\def \beq{\begin{equation}}
\def \bo{B^0}
\def \cpp{C_{\pi \pi}}
\def \eeq{\end{equation}}
\def \lpp{\lambda_{\pi \pi}}
\def \ob{\overline{B}^0}
\def \rpp{R_{\pi \pi}}
\def \spp{S_{\pi \pi}}
\def \beqa{\begin{eqnarray}}
\def \eeqa{\end{eqnarray}}
%%%%%%%%%%%%%%%%%%%%%%%%%%%%%%%%%%%%%%%%%%%%%%%%%%%%
\begin{center}
\vspace{2.0cm}
\end{center}

\section{Introduction}

Understanding nonleptonic $B$ meson decays is crucial for testing
the standard model, and also for uncovering the trace of new
physics. The simplest case is two-body nonleptonic $B$ meson
decays, for which Bauer, Stech and Wirbel proposed the
factorization assumption (FA) in their pioneering work \cite{BSW}.
Considerable progress, including the generalized FA 
\cite{Cheng94,Cheng96,Soares} and
QCD-improved FA (QCDF) \cite{BBNS}, has been made since this proposal. On
the other hand, techniques to analyze hard exclusive hadronic
scattering has been developed by Brodsky and Lepage \cite{LB} based on
collinear factorization theorem in perturbative QCD (PQCD). A
modified framework based on $k_T$ factorization theorem has been
given in \cite{BS,LS}, and extended to exclusive $B$ meson decays
in \cite{LY1,CL,YL,CLY}. The infrared finiteness and gauge
invariance of $k_T$ factorization theorem was shown explicitly in
\cite{NL}. Using the PQCD approach, we have
investigated  the dynamics of nonleptonic $B$ meson decays
\cite{KLS,LUY,KS}. Our observations are summarized as follows:
\begin{enumerate}
\item FA is approximately correct, as our computation shows that
nonfactorizable contributions in charmless $B$ meson decays are
negligible.

\item Penguin amplitudes are enhanced, as the PQCD formalism
inludes dynamics from the energy region as low as 
$\sqrt{\bar\Lambda m_b}$. Here $\bar\Lambda\equiv m_B-m_b$, 
being the $B$ meson and $b$ quark mass difference.
Note that $\sqrt{\bar\Lambda m_b}$ is much lower than $m_b$(or $m_b/2$)
which is often used as the energy scale.

\item Annihilation diagrams give rise to large short-distance
strong phases through $(S+P)(S-P)$ penguin operators.

\item The sign and magnitude of CP asymmetries in
two-body nonleptonic $B$ meson decays can be calculated, and we
have predicted relatively large CP asymmetries in the $B\to
K^{(*)}\pi$
\cite{KLS} and $\pi\pi$ modes\cite{LUY,KS}.
\end{enumerate}

In this talk we summarize the PQCD method and
important theoretical issues, and 
describe the analysis of branching ratios of B-meson
decays including $B \to D^{(*)} \pi$ decays and discuss the origin
of large direct CP-violation in $B \to \pi\pi$ process.  
   
\section{Ingredients of PQCD and Theoretical Issues}

{\bf End Point Singularity and Form Factors:} 
If we calculate the $B\to\pi$ transition form factor $F^{B\pi}$ 
at large recoil using
the Brodsky-Lepage formalism \cite{Bro,BSH}, a difficulty immediately
occurs. The lowest-order diagram for the hard amplitude is proportional to 
$1/(x_1 x_3^2)$, $x_1$ being the momentum fraction associated with the
spectator quark on the $B$ meson side and $x_3$ the momentum fraction 
carried by the spectator quark in final pion. 
If the pion distribution amplitude
vanishes like $x_3$ as $x_3\to 0$ (in the leading-twist, {\it i.e.},
twist-2 case), $F^{B\pi}$ is logarithmically divergent. If the pion
distribution amplitude is a constant as $x_3\to 0$ (in the
next-to-leading-twist, {\it i.e.}, twist-3 case), $F^{B\pi}$ even becomes
linearly divergent. These end-point singularities have also appeared in
the evaluation of the nonfactorizable and annihilation amplitudes in QCDF.

Note that in the above discussion parton transverse momenta
$k_{\perp}$ has been neglected. More accurately, we have
\begin{equation}
{1 \over x_1\,\, x_3^2 M_B^4} \,\, \rightarrow
\,\, {1 \over (x_3\, M_B^2 + k_{3\perp}^2) \,\,
[x_1x_3\, M_B^2 + (k_{1\perp} - k_{3\perp})^2]}
\label{eq:4} 
\end{equation}
and the end-point singularity is smeared out. 
More precise analysis including the Sudakov and threshold resummation
effects
has been given \cite{KLS}.

In PQCD, we can calculate the form factors for $B \to P,V$ transitions
\cite{CKL:a,Kurimoto}.

\vspace{5mm}
{\bf Dynamical Penguin Enhancement vs Chiral Enhancement:} 
The typical hard scale is about 1.5 GeV as discussed in Ref.\cite{KLS}.
Since the RG evolution of the Wilson coefficients $C_{4,6}(t)$ increase
drastically as $t < M_B/2$, while that of $C_{1,2}(t)$ remain almost
constant, we can get a large enhancement effects from both Wilson
coefficents and matrix elements in PQCD. 
 
In general the amplitude can be expressed as
\begin{equation}
Amp \sim [a_{1,2} \,\, \pm \,\, a_4 \,\,
\pm \,\, m_0^{P,V}(\mu) a_6] \,\, \cdot \,\, <K\pi|O|B>
\label{eq:2}
\end{equation}
with the chiral factors $m_0^P(\mu)=m_P^2/[m_1(\mu)+m_2(\mu)]$ for
pseudoscalar meson 
and $m_0^{V}= m_V$ for vector meson.
To accommodate the $B\to K\pi$ data in FA and
QCD-factorization approach, one relies on the chiral enhancement by
increasing  $m_0$ to as large as about 3 GeV at $\mu=m_b$ scale.
So two methods accomodate large branching ratios of $B \to K\pi$.
But, there is no such adjustable parameter for $B \to PV$ decay.
For $B \to PV$ there is no chiral factor in LCDAs of the vector meson.
Here $m_0=m_V$. It is difficult to explain large penguin contribution
in QCDF\cite{CK}. 

We can test whether a dynamical enhancement 
or a chiral enhancement is responsible
for the large $B \to K\pi$ branching ratios 
by measuring the $B \to VP,VV$ modes.
In these modes penguin contributions dominate, 
such that their branching ratios are
insensitive to the variation of the unitarity angle $\phi_3$.
Our prediction for various modes are shown at Table 2. 
We point out that QCDF can not globally fit the
experimental data for $B\to PP,VP$ and $VV$ modes simultaneously
with same sets of free parameters $(\rho_H,\phi_H)$ and $(\rho_A,\phi_A)$
\cite{zhu}.

%%%%%%%%%%%%%%%%%%%%%%% Branching ratios (1) %%%%%%%%%%%%%%%%%%%%
\begin{table*}[h!]
\caption{Branching ratios of $B \to \pi\pi, K\pi$and $KK$ decays 
with $\phi_3=80^0$, $R_b=\sqrt{\rho^2+\eta^2}=0.38$. 
Here we adopted $m_0^{\pi}=1.3$ GeV,
$m_0^{K}=1.7$ GeV and $0.36<\omega_B<0.44$.
Unit is $10^{-6}$.} 
\label{TABLE1 }
\begin{center}
\begin{tabular} {|c|ccc|c|c|} \hline 
Decay Channel & CLEO & BELLE & BABAR & ~~~World Av.~~~ & ~~~~~~~PQCD~~~~~~~  \\
\hline  
$\pi^{+}\pi^{-}$ & $4.5^{+1.4+0.5}_{-1.2-0.4}$ &
 $4.4\pm 0.6 \pm 0.3$ &
 $4.7\pm 0.6 \pm 0.2$ &  
$4.6 \pm 0.4$ &
$5.93-10.99$  \\
$\pi^{+}\pi^{0}$ & $4.6^{+1.8+0.6}_{-1.6-0.7}$ & 
 $5.3 \pm 1.3 \pm0.5$ &
 $5.5^{+1.0}_{-0.9} \pm 0.6$ & $5.3\pm0.8$ & 
$2.72-4.79$    \\ 
$\pi^{0}\pi^{0}$ & $<4.4$ & 
 $<4.4$ &  $<3.6$ & $<3.6$ &
  $0.33-0.65$    \\ 
\hline
$K^{\pm}\pi^{\mp}$ &  
 $18.0^{+2.3+1.2}_{-2.1-0.9}$ &
 $18.5 \pm1.0 \pm0.7$ &  
 $17.9\pm 0.9 \pm 0.7$ &
 $18.2 \pm 0.8$ &  
 $12.67-19.30$    \\ 
$K^{0}\pi^{\mp}$ & 
 $18.8^{+3.7+2.1}_{-3.3-1.8}$ &
 $22.0 \pm1.9 \pm1.1$  &   
 $20.0\pm 1.6 \pm 1.0$ &
 $20.6 \pm 1.4$ &  
 $14.43-26.26$    \\ 
$K^{\pm}\pi^{0}$ &
 $12.9^{+2.4+1.2}_{-2.2-1.1}$ &
 $12.8 \pm1.4^{+1.4}_{-1.0}$ &  
 $12.8^{+1.2}_{-1.0}\pm 1.0$ &
 $12.8 \pm 1.1$ &  
  $7.87-14.21$    \\
$K^{0}\pi^{0}$ &
 $12.8^{+4.0+1.7}_{-3.3-1.4}$ &
 $12.6 \pm2.4 \pm1.4$ &  
 $10.4 \pm 1.5 \pm 1.8$ &
 $11.5 \pm 1.7$ & 
 $7.92-14.27$    \\ 
\hline 
$K^{\pm}K^{\mp}$ &
 $<0.8$ &
 $<0.7$ &  
 $<0.6$ &
 $<0.6$ & 
 $0.06$    \\ 
$K^{\pm}\bar{K}^{0}$ &
 $<3.3$ &
 $<3.4$ &  
 $<2.2$ &
 $<2.2$ & 
 $1.4$    \\ 
$K^{0}\bar{K}^{0}$ &
 $<3.3$ &
 $<3.2$ &  
 $<1.6$ &
 $<1.6$ & 
 $1.4$    \\ 
\hline
\end{tabular}
\end{center}
\end{table*} 
%%%%%%%%%%%%%%%%%%%%%%%%%%%%%%%%%%%%%%%%%%%%%%%%%%%%%%%%%%%%%%%%%%

\vspace{5mm}
{\bf Fat Imaginary Penguin in Annihilation:} 
There is a folklore that annihilation contribution is negligible
compared to the W-emission. For this reason, annihilation contribution
was not included in the general factorization approach and the first
paper on QCD-factorization by Beneke et al. \cite{BBNS:99}.
In fact there is a suppression effect for the operators with structure
$(V-A)(V-A)$ because of a mechanism similar to the helicity
suppression for $\pi \to \mu \nu_{\mu}$. However annihilation from 
the operators $O_{5,6,7,8}$ with the structure $(S-P)(S+P)$ via Fiertz
transformation survive under the helicity suppression. 
Moreover, they provide large imaginary part. 
The real part of the factorized annihilation contribution
becomes small because there is a cancellation between left-handed
gluon exchanged diagram and right-handed gluon exchanged one as shown in
Table 1 of ref.\cite{KS}. 
This mostly pure imaginary value of annihilation is a main
source of large CP asymmetry in $\pi^{+}\pi^{-}$ and $K^{+}\pi^{-}$ decays.
In Table 3 we summarize the CP asymmetry in 
$B \to K(\pi)\pi$ decays with experimental measurements.
%%%%%%%%%%%%%%%%%%%%%%%%%%%%%%%%%%%%%%%%%%%%%%%%%%%%%%%%%%%%%%%%%%%%%%%%%

\vspace{5mm}
{\bf Small Strong Phase for FA and QCDF:} 
We have seen that the dominant strong phase in PQCD
comes from the factorizable annihilation
diagram\cite{KLS}. For FA and QCDF, stong phases
come from the Bander-Silverman-Soni (BSS) mechanism\cite{BSS}
and from the final state interaction (FSI). In fact,
the two sources of strong phases in the FA
and QCDF are strongly suppressed by the charm mass
threshold and by the end-point behavior of meson wave functions.
So the strong phase in these approaches is almost zero 
without soft-annihilation contributions.

%%%%%%%%%%%%%%%%% Branching ratio (2)  %%%%%%%%%%%%%%%%%%%%%%%%%%%
\begin{table*}[thb]
\caption{Branching ratios of $B \to \phi K^{(*)}$and $K^{*}\pi$ decays 
with $\phi_3=80^0$, $R_b=\sqrt{\rho^2+\eta^2}=0.38$. 
Here we adopted $m_0^{\pi}=1.3$ GeV
and $m_0^{K}=1.7$ GeV.
Unit is $10^{-6}$.} 
\label{TABLE2}
\begin{center}
\begin{tabular}{|c|ccc|c|c|} \hline
Decay Channel & CLEO & BELLE & BABAR 
& ~~~~~~ World Av.~~~~~~ &~~~~~~~~PQCD~~~~~~~~   \\
\hline  
$\phi K^{\pm}$ & 
 $5.5^{+2.1}_{-1.8}\pm 0.6$ &
 $9.4 \pm 1.1 \pm 0.7$ &  
 $10.0^{+0.9}_{-0.8}\pm 0.5$ &   
 $9.3 \pm 0.8$ &
 $8.1-14.1$  \\
$\phi K^{0}$ & 
 $ 5.4^{+3.7}_{-2.7} \pm 0.7 $ &
 $9.0 \pm 2.2 \pm 0.7$ &  
 $7.6^{+1.3}_{-1.2}\pm 0.5 $ &
 $7.7 \pm 1.1$  &
 $7.6-13.3$    \\ 
\hline
$\phi K^{*\pm}$ & 
 $10.6^{+6.4+1.8}_{-4.9-1.6}$ &  
 $6.7^{+2.1+0.7}_{-1.9-1.0} $ & 
 $12.1^{+2.1}_{-1.9} \pm 1.1$  &
 $9.4 \pm 1.6$ &
 $12.6-21.2$ \\
$\phi K^{*0}$ & 
 $11.5^{+4.5+1.8}_{-3.7-1.7} $ &
 $10.0^{+1.6+0.7}_{-1.5-0.8} $ &  
 $11.1^{+1.3}_{-1.2}\pm 0.8 $ &
 $10.7 \pm 1.1$    &
 $11.5-19.8$    \\ 
\hline
$K^{*0} \pi^{\pm}$ & 
 $7.6^{+3.5}_{-3.0} \pm 1.6$ &
 $19.4^{+4.2+4.1}_{-3.9-7.1}$ &  
 $15.5 \pm 3.4 \pm 1.8$ &
 $12.3 \pm 2.6$   & 
 $10.2-14.6$  \\
$K^{*\pm}\pi^{\mp}$ & 
 $16^{+6}_{-5} \pm 2 $ &
 $<30$ &  
 $-$ &
 $16 \pm 6$    &
 $8.0-11.6$    \\
$K^{*+} \pi^{0}$ & 
 $<31$ &
 $-$ &  
 $-$ &
 $<31$   & 
 $2.0-5.1 $  \\
$K^{*0}\pi^{0}$ & 
 $<3.6 $ &
 $<7$ &  
 $-$ &
 $<3.6$    &
 $1.8-4.4$  \\   
\hline 
\end{tabular}
\end{center}
\end{table*} 
%%%%%%%%%%%%%%%%%%%%%%%%%%%%%%%%%%%%%%%%%%%%%%%%%%%%%%%%%%%%%%%%%%%%%%%%%

\section{Numerical Results in Charmless B-decays}

{\bf Branching ratios in $B \to PP,VP$ and $VV$:} 
The PQCD approach allows us to calculate 
the amplitudes for charmless B-meson decays
in terms of light-cone distribution amplitudes upto twist-3. 
We focus on decays
whose branching ratios have already been measured. 
We take allowed ranges of shape parameter for the B-meson wave function as 
$\omega_B = 0.36-0.44$, which accommodate reasonable form factors: 
both $F^{B\pi}(0)=0.27-0.33$ and $F^{BK}(0)=0.31-0.40$. 
We use values of chiral factor
with $m_0^{\pi}=1.3 GeV$ and $m_0^{K}=1.7 GeV$.
Finally we obtain branching ratios for $B\to K(\pi)\pi$ 
\cite{KLS,LUY}, 
$K\phi$ \cite{CKL:a,Mishima} $K^{*}\phi$\cite{CKL:b} and 
$K^{*}\pi$\cite{KLS},
which are in agreement with experimental data. In table 1 and 2
we summarize our predictions for various decays with updated
experimental measurements which was reported 
at FPCP-2003 workshop\cite{Bona}.

%%%%%%%%%%%%%%%%%%%%%%% CP-Asymmety in Kpi, pipi decays  %%%%%%%%%%%%%%%%%%%%
\begin{table*}[t!]
\caption{Direct CP-asymmetry in $B \to K \pi, \pi\pi $ decays 
with $\phi_3=40^0 \sim 90^0$, $R_b=\sqrt{\rho^2+\eta^2}=0.38$. 
Here we adopted $m_0^{\pi}=1.3$ GeV and $m_0^{K}=1.7$ GeV for the PQCD
results.}
\begin{tabular}{|c||c|c||c|c|c|} \hline 
~~Direct~~$A_{CP}(\%)$~~~~~~& ~~~~~BELLE~~~~~   & ~~~~~BABAR~~~~~   
& ~~~~~~~~~~PQCD~~~~~~~~~~ & ~~~~~~~QCDF~~~~~~ & Charming Penguin ($|A_{cp}|$)\\ \hline 
$\pi^{+}\pi^{-}$ & $77\pm27\pm8$ & $30 \pm 25 \pm 4$ & 
 $16.0 \sim 30.0$ & $-6\pm12$ & $39\pm20$  \\ \hline
$\pi^{+}\pi^{0}$ & $-14\pm24^{+5}_{-4}$ & $-3 \pm 18 \pm 2$ & 
 $0.0$ & 0.0  & $0.0$  \\ \hline \hline
$\pi^{+} K^{-}$ & $-7 \pm 6 \pm 1$ & $-10.2\pm5.0\pm1.6$ & 
$-12.9 \sim -21.9  $ & $5\pm9$ & $21\pm 12$  \\ \hline
$\pi^{0}K^{-}$ & $23\pm11^{+1}_{-4}$& $-9.0 \pm 9.0 \pm1.0$ & 
 $-10.0 \sim -17.3$ & $7\pm9$ & $22\pm13$  \\ \hline
$\pi^{-}\bar{K}^{0}$ & $7^{+9+1}_{-8-3}$ & $-4.7 \pm 13.9$ & 
 $-0.6 \sim -1.5$ & $1\pm1$  & $0.0$   \\ \hline 
\end{tabular}
\label{TABLE3}
\end{table*} 

%%%%%%%%%%%%%%%%%%%%%%%%%%%%%%%%%%%%%%%%%%%%%%%%%%%%%%%%%%%%%%%%%%%%%%%%%

\vspace{5mm}
{\bf CP Asymmetry of $B \to \pi\pi, K\pi$:}
Because we have a large imaginary contribution from factorized 
annihilation diagrams,
we predict large CP asymmetry ($\sim 25 \%$) 
in $B^0 \to \pi^{+}\pi^{-}$ decay
and about $-15 \%$ CP violation effects in  $B^0 \to K^{+}\pi^{-}$.
The detail prediction is given in Table 3.
The precise measurement of direct CP asymmetry (both magnitude and sign) 
is a crucial way to test factorization models 
which have different sources of strong phases.

\vspace{5mm}
{\bf Understanding of $Br(K^{*}\pi)$ and $Br(\omega K^+/\omega \pi^+)$:}
In PQCD, penguin contributions enhances both $Br(K\pi)$ and
$Br(K^*\pi)$.
As noted before, in FA and QCDF, the penguin enhancement can achieved
by taking $m_0$ as large as 3 $GeV$, however, they fail to explain 
the large branching ratio for $K^{*}\pi$ decay. 
 
Another hot issue is how can we understand
$Br(\omega K/\omega \pi) \sim 1$. Since $\omega K^+$ decay is
penguin dominant process and $\omega \pi^+$ decay is tree dominant
one: $(a_1 + x a_2)$, it is hard to get the same branching ratio.
PQCD method predicts $Br(\omega K^+) \sim 3.22 \times 10^{-6}$ and
$Br(\omega \pi^+) \sim 6.20 \times 10^{-6}$, which 
still has about factor 2 differences between them.
We need more precise measurements on $\omega K^{+}$ decay. 

%%%%%%%%%%%%%%%%%%%%%%%%%%%%
\begin{figure*}[t!]
\begin{center}
\resizebox{25pc}{!} {\includegraphics[angle=-90,width=15.0cm]{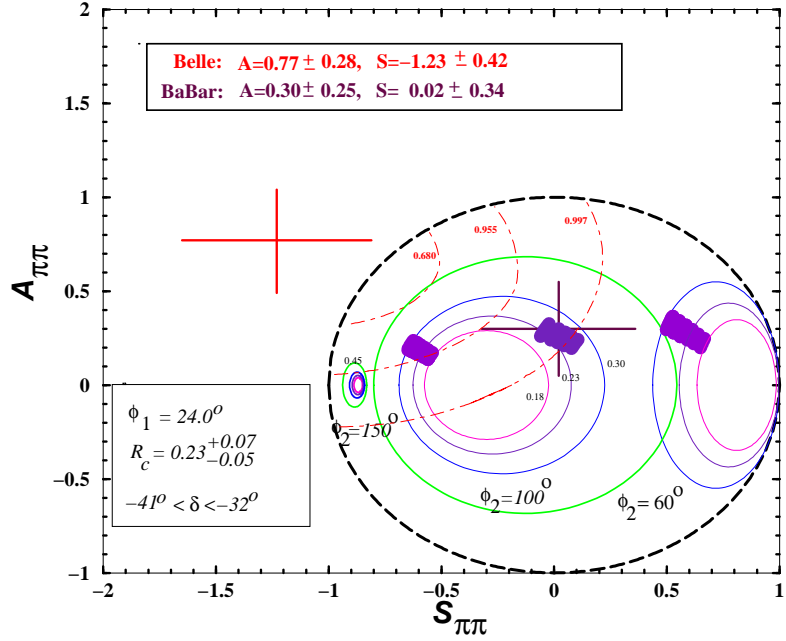}} 
\caption{Plot of $A_{\pi\pi}$ versus $S_{\pi\pi}$  for various values
of $\phi_2$ with $\phi_1=24.3^o$, $0.18 < R_c < 0.30$ and $-41^o <
\delta < -32^o$ in the pQCD method.}
\end{center}
\label{fig:cpipi}
\end{figure*}
%%%%%%%%%%%%%%%%%%%%%%%%%%%
%%%%%%% Determination of Phi_2 and Phi_3 from B to pipi/Kpi %%%%%%%%%%%%%
\section{Extraction of $\phi_2$ from $B \to \pi^{+}\pi^{-}$}
Even though isospin analysis of $B \to \pi\pi$ can provide a clean way
to determine $\phi_2$, it might be difficult in practice because of
the small branching ratio of $B^0 \to \pi^0\pi^0$.
Here we describe the time-dependent analysis of $B^0(t) \to \pi^{+}\pi^{-}$.
Since penguin contributions are sizable 
and are about 20-30 \% of the total amplitude,
we expect that direct CP violation can be large if strong phases are different
in the tree and penguin diagrams.

The ratio between penguin and tree amplitudes is $R_c=|P_c/T_c|$
(here we use the c-convention notation) 
and the strong phase difference
between penguin and tree amplitudes $\delta=\delta_P-\delta_T$.
The time-dependent asymmetry measurement provides two equations for
$C_{\pi\pi}$ and $S_{\pi\pi}$ in terms of three unknown variables 
$R_c,\delta$ and $\phi_2$\cite{GR}.

Since pQCD provides us $R_c=0.23^{+0.07}_{-0.05}$ and $-41^o
<\delta<-32^o$, the allowed range of $\phi_2$ at present stage is
determined as $55^o <\phi_2< 100^o$ as shown in Figure \ref{fig:cpipi}. 

According to the power counting rule in the pQCD approach,
the factorizable annihilation contribution with large imaginary part
becomes big and give a negative strong phase from 
$-i\pi\delta(k_{\perp}^2-x\,M_B^2)$.
Therefore we have a relatively large
strong phase in contrast to QCD-factorization ($\delta\sim 0^o$) 
and predict large direct CP violation effect 
in $B^0\to \pi^{+}\pi^{-}$ 
with $A_{cp}(B^0 \to \pi^{+}\pi^{-}) = (23\pm7) \%$, 
which will be tested by more precise experimental measurement 
within two years. 

In the numerical analysis, since the data by Belle
collaboration\cite{belle} 
is located outside allowed physical regions, we considered the
recent BaBar measurement\cite{babar} with $90\%$ C.L. interval
taking into account the systematic errors:
\begin{itemize}
\item[$\bullet$]
$S_{\pi\pi}=  0.02\pm0.34\pm0.05$ 
\hspace{3mm} [-0.54,\hspace{2mm} +0.58]
\item[$\bullet$]
$A_{\pi\pi}= 0.30\pm0.25\pm0.04$ 
\hspace{3mm} [-0.72,\hspace{2mm} +0.12].
\end{itemize}
The central point of BaBar data corresponds to $\phi_1 = 24^o$,
$\phi_2 = 78^o$, and $\phi_3 = 78^o$ when the Standard Model works.   

Even though the data by Belle
collaboration\cite{belle} 
is located outside allowed physical regions, we can have overlapped
ranges within 2 $\sigma$ bounds. 

%%%%%%%%%%%%%%%%%%%%%%%%%%%%%%%%%%%%%%%%%%%%%%%%%%%%%%%%%%%%%%%%%
%%%%%%%%%%%%%%%%%%%%%%%%%%%%
\begin{figure*}[t!]
\begin{center}
\resizebox{25pc}{!} {\includegraphics[angle=-90,width=15.0cm]{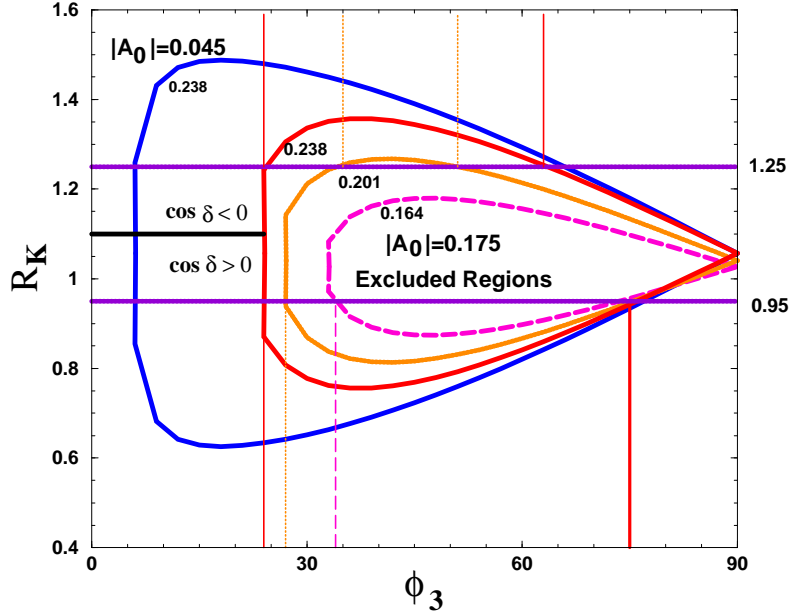}} 
\caption{Plot of $R_K$ versus $\phi_3$ with $r_K=0.164,0.201$ and $0.238$.}
\end{center}
\label{fig:RK}
\end{figure*}
%%%%%%%%%%%%%%%%%%%%%%%%%%%

\section{Extraction of $\phi_3$ from $B \to K\pi$}

By using tree-penguin interference in $B^0\to K^{+}\pi^{-}(\sim
T^{'}+P^{'})$ versus $B^{+}\to K^0\pi^{+}(\sim P^{'})$, CP-averaged
$B\to K\pi$ branching fraction may lead to non-trivial constaints
on the $\phi_3$ angle\cite{fle-man}. In order to determine $\phi_3$,
we need one more useful information 
on CP-violating rate differences\cite{gr-rs02}.
Let's introduce the following observables :
\beqa
R_K &=&{\overline{Br}(B^0\to K^{+}\pi^{-}) \,\, \tau_{+} \over
\overline{Br}(B^+\to K^{0}\pi^{+}) \,\, \tau_{0} }
= 1 -2\,\, r_K \, cos\delta \, \, cos\phi_3 + r_K^2 \nonumber \\
\cr
A_0 &=&{\Gamma(\bar{B}^0 \to K^{-}\pi^{+}) - \Gamma(B^0 \to
K^{+}\pi^{-}) \over \Gamma(B^{-}\to \bar{K}^0\pi^{-}) +
 \Gamma(B^{+}\to \bar{K}^0\pi^{+}) } \nonumber \\
&=& A_{cp}(B^0 \to K^{+}\pi^{-}) \,\, R_K = -2 r_K \, sin\phi_3 \,sin\delta.
\eeqa
where $r_K = |T^{'}/P^{'}|$ is the ratio of tree to penguin amplitudes
in $B\to K\pi$
and $\delta = \delta_{T'} -\delta_{P'}$ is the strong phase difference
between tree and penguin amplitides.
After eliminating $sin\delta$ in Eq.(8)-(9), we have
\beq
R_K = 1 + r_K^2 \pm \sqrt{(4 r_K^2 cos^2\phi_3 -A_0^2 cot^2\phi_3)}.
\eeq
Here we obtain $r_K = 0.201\pm 0.037$ from the PQCD analysis 
and $A_0=-0.11\pm 0.065$ by combining recent BaBar
measurement on CP asymmetry of $B^0\to K^+\pi^-$: 
$A_{cp}(B^0\to K^+\pi^-)=-10.2\pm5.0\pm1.6 \%$ \cite{babar}
with present world averaged value of  $R_K=1.10\pm 0.15$\cite{rk}.

PQCD method provides $\delta_{P'} = 157^o$, $\delta_{T'} = 1.4^o$
and the negative $cos\delta$: $cos\delta= -0.91$.
As shown in Fig.2, we can constrain the allowed range of $\phi_3$ 
within $1\,\sigma$ range of World Averaged $R_K$ as follows:
\begin{itemize}
\item[$\bullet$]For $cos\delta < 0$, $r_K=0.164$: we can exclude
$0^o \leq \phi_3 \leq 6^0$. 
\item[$\bullet$]For $cos\delta < 0$, $r_K=0.201$: we can exclude
$0^o \leq \phi_3 \leq 6^0$ and $ 35^o \leq \phi_3 \leq 51^0$. 
\item[$\bullet$]For $cos\delta < 0$, $r_K=0.238$: we can exclude
$0^o \leq \phi_3 \leq 6^0$ and $ 24^o \leq \phi_3 \leq 62^0$.
\end{itemize}
When we take the central value of $r_K=0.201$,
$\phi_3$ is allowed within the ranges of 
$51^o \leq \phi_3 \leq 90^o$, which is consistent with 
the results by the model-independent 
CKM-fit in the $(\rho,\eta)$ plane.

%%%%%%%%%%%%%%%%%%%%%%%%%%%%%%%%%%%%%%%%%%%%%%%%%%%%%%%%%%%%%%%%%%

\section{Large Nonfactorizable contribution in Charmed B-decays}
{\bf Large $a_2/a_1$ in $B \to D^{(*)}\pi$ Decays:} 
Being free from the end-point singularities, all topologies of
decay amplitudes of charmed decays can be computed in the PQCD approach,
including the nonfactorizable color-suppressed diagram.
This amplitude can not be calculated in the QCDF approach based
on collinear factorization theorem because of the existence of the
singularities. However, we found that this amplitude is crucial
for explaining the observed $B\to D^{(*)}\pi$ branching ratios,
since it is not suppressed by the Wilson coefficient (proportional
to $C_2/N_c$), and provides a large strong phase required by the
isospin relation. The tree annihilation amplitude, also
contributing to the strong phase, is not important.
As stated above, we have predicted large
strong phases from the scalar-penguin annihilation amplitudes,
which are required by the large CP asymmetries in two-body
charmless decays. The success in predicting the strong phases from
the nonfactorizable color-suppressed amplitudes for the two-body
charmed decays further supports $k_T$ factorization theorem.
The conclusion is that the short-distance
strong phase is sufficient to account for the $B\to D^{(*)}\pi$
data. A long-distance strong phase from final-state interaction
may be small, though it should exist.
Finally we obtained $a_2/a_1 \sim 0.4-0.5$ and $Arg(a_2/a_1)\sim -42^o$
by including annihilation contributions\cite{KLKLS}.
%%%%%%%%%%%%%%%%%%%%%%%%%%%%%%%%%%%%%%%%%%%%

\section{Summary and Outlook}
In this talk we have discussed ingredients of PQCD approach 
and some important theoretical issues with numerical results.
The PQCD factorization approach provides a useful theoretical framework
for a systematic analysis on non-leptonic two-body B-meson decays. 
Specially pQCD predicted large direct CP asymmetries
in $B^0 \to \pi^{+}\pi^{-}, K^{+}\pi^{-}$, $K^{*\pm}\pi^{\mp}$ and
$K^{*\pm}\pi^{0}$ decays, 
which will be a crucial touch stone
to distinguish PQCD approach from others 
in future precise measurement.

We discussed two methods to determine weak phases 
$\phi_2$ and $\phi_3$ within the pQCD approach 
through 1) Time-dependent asymmetries in $B^0\to
\pi^{+}\pi^{-}$, 2) $B\to K\pi$ processes via penguin-tree
interference. We can get interesting bounds on $\phi_2$
and $\phi_3$ from present experimental measurements.
From BaBar measurement of CP asymmetry 
in $\pi^{+}\pi^{-}$ decay the  prefered CKM weak phases are:
$\phi_1=(24\pm 2)^{o}$, $\phi_2=(78\pm 22)^{o}$ 
and $\phi_3=(78\pm 22)^{o}$.

\begin{center}{\bf\Large Acknowledgments}
\end{center}

We wish to acknowlege the fruitful collaboration with H.-N. Li.
YYK thanks H.~Sagawa for statistical bounds for Belle data
and M. Bona for updated experimental results.
This work was supported in part by the Japan Society for
the Promotion of Science, 
and in part by Ministry of Education, Science and Culture, Japan.

% choose bibtex style depending on layout style and options used in
% sample:

\end{document}